\documentclass[fleqn,10pt]{wlscirep}
\usepackage[utf8]{inputenc}
\usepackage[T1]{fontenc}
\title{Design guidelines for the SPICE parameters of waveform-selective metasurfaces varying with the incident pulse width at a constant oscillation frequency}

\author[1]{Shiori Imai}
\author[2]{Haruki Homma}
\author[2]{Kairi Takimoto}
\author[2]{Mizuki Tanikawa}
\author[3]{Jin Nakamura}
\author[3]{Masaya Kaneko}
\author[3]{Yuya Osaki}
\author[4]{Kiichi Niitsu}
\author[5]{Cheng Yongzhi}
\author[2]{Ashif Aminulloh Fathnan}
\author[2,*]{Hiroki Wakatsuchi}
\affil[1]{Department of Electrical and Mechanical Engineering, Faculty of Engineering, Nagoya Institute of Technology, Nagoya, Aichi 466‑8555, Japan}
\affil[2]{Department of Engineering, Graduate School of Engineering, Nagoya Institute of Technology, Nagoya, Aichi 466‑8555, Japan}
\affil[3]{Kyoto Engineering Center, Meitec Corporation, Kyoto, 600-8216, Japan}
\affil[4]{Graduate School of Informatics, Kyoto University, Kyoto, 606-8501, Japan}
\affil[5]{School of Information Science and Engineering, Wuhan University of Science and Technology, Wuhan, 430081, China}

\affil[*]{wakatsuchi.hiroki@nitech.ac.jp}

\begin{abstract}
In this study, we numerically demonstrate how the response of recently reported circuit-based metasurfaces is characterized by their circuit parameters. These metasurfaces, which include a set of four diodes as a full wave rectifier, are capable of sensing different waves even at the same frequency in response to the incident waveform, or more specifically the pulse width. This study reveals the relationship between the electromagnetic response of such waveform-selective metasurfaces and the SPICE parameters of the diodes used. First, we show that reducing a parasitic capacitive component of the diodes is important for realization of waveform-selective metasurfaces in a higher frequency regime. Second, we report that the operating power level is closely related to the saturation current and the breakdown voltage of the diodes. Moreover, the operating power range is found to be broadened by introducing an additional resistor into the inside of the diode bridge. Our study is expected to provide design guidelines for circuit-based waveform-selective metasurfaces to select/fabricate optimal diodes and enhance the waveform-selective performance at the target frequency and power level.
\end{abstract}
\begin{document}

\flushbottom
\maketitle
%
%
\thispagestyle{empty}

\section*{Introduction}

Artificially engineered structures referred to as metamaterials and metasurfaces enable ready tailoring of electromagnetic fields at will \cite{MTMbookEngheta,smithDNG1D,EBGdevelopment,yu2011light,yu2014flat}. Compared to conventional materials that respond to an incident wave in accordance with the reaction of composite molecules, the response of metasurfaces is characterized by composite unit cells that are much larger than molecules but still at the subwavelength scale and can thus be arbitrarily customized to achieve a range of electromagnetic properties including negative refraction \cite{smithDNG2D1} and a large impedance surface \cite{EBGdevelopment}. In addition, these metasurfaces with exotic properties can be exploited to develop applied devices or systems such as invisibility cloaks \cite{pendryCloaking,schurig2006metamaterial}, diffraction-limit-breaking lenses \cite{pendryperfetLenses,fangSuperlens}, perfect absorbers \cite{mtmAbsPRLpadilla,ultraThinAbs,My1stAbsPaper,watts2012metamaterial}, antennas \cite{ziolkowski2011metamaterial,barbuto2021metasurfaces}, analogue computation systems \cite{silva2014performing,mohammadi2019inverse,zangeneh2021analogue} and beamforming systems \cite{sievenpiper2003two,yu2011light,pfeiffer2013metamaterial,yu2014flat,shaltout2019spatiotemporal} as intelligent reflecting surfaces (IRSs) \cite{wu2019towards,di2020smart,gradoni2021smart}. 

In particular, these artificially constructed materials attain a higher performance level by including nonlinearity, which offers capabilities that are not available with a simple combination of linear media or structures \cite{THzActiveMTMpadilla,kivshar2003nonlinear,lapine2014colloquium,li2017nonlinear}. For instance, nonlinear metasurface-based absorbers dissipate the energy of a strong electromagnetic wave to prevent electromagnetic interference issues while permitting propagation of a small signal for antenna communication even at the same frequency \cite{aplNonlinearMetasurface,kim2016switchable,li2017high,zhou2021high}. Additionally, the electromagnetic response of IRSs is altered in accordance with surrounding electromagnetic fields or incoming communication signals, which is achieved by including nonlinear circuit elements together with a field-programmable gate array (FPGA) and an external direct current (DC) source \cite{zhang2018space,zhang2021wireless}. 

Moreover, a series of recent studies demonstrated that metasurfaces composed of Schottky diodes were capable of sensing different waves even at the same frequency in response to the incoming waveform, or more specifically the pulse width \cite{wakatsuchi2013waveform,eleftheriades2014electronics,wakatsuchi2015waveformSciRep,wakatsuchi2019waveform}. This waveform selectivity provides a new degree of freedom to control electromagnetic waves at a single frequency, which has thus far been exploited to mitigate electromagnetic interference \cite{wakatsuchi2019waveform}, design antennas \cite{vellucci2019waveform,barbuto2020waveguide,ushikoshi2022pulse} and related microwave devices \cite{fathnan2022method,tashirometasurface} and control communication signals \cite{wakatsuchi2015waveformSciRep,ushikoshi2019experimental,f2020temporal}. Although the electromagnetic response related to the conducting geometry is well known to be improved by existing methods, including the use of equivalent circuit models \cite{baena2005equivalent,MyCWeqCircuitPaper}, the relationship between the waveform-selective performance and SPICE parameters of diodes remains unclear, which is important for the design of waveform-selective metasurfaces. For this reason, this study reveals how the waveform selectivity is determined by the SPICE parameters used for waveform-selective metasurfaces. In particular, we present simulation results to show how the SPICE parameters are associated with the operating frequency and power of waveform-selective metasurfaces. Thus, our study provides design guidelines for circuit-based waveform-selective metasurfaces to select/fabricate optimal diodes and enhance the waveform-selective performance at the target frequency and power level.

\section*{Results}

\begin{figure}[tb]
\centering
\includegraphics[width=0.7\linewidth]{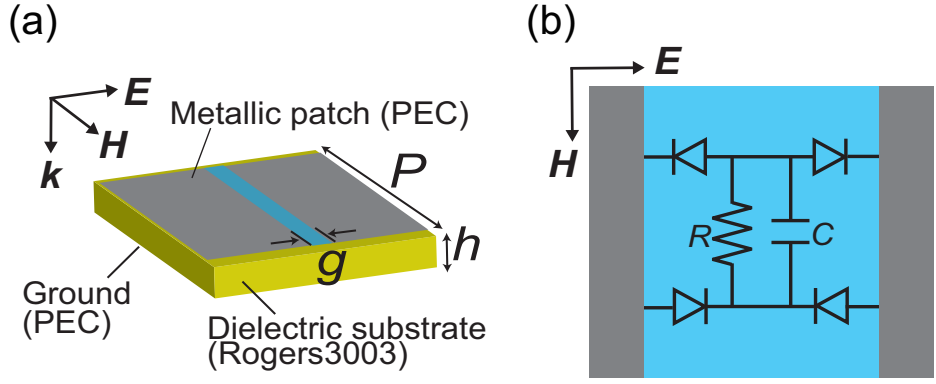}
\caption{Simulated periodic unit cell model of the waveform-selective metasurface. (a) Unit cell and (b) circuit deployed between metallic patches. Periodic boundaries were applied in the incident electric and magnetic field directions. }
\label{fig:1}
\end{figure}

\begin{table}[tb]
\centering
\begin{tabular}{|c|c|}
\hline
Parameter & Value \\
\hline
$I_b$ & 1.00$\times 10^{-5}$ A \\
\hline
$I_S$ & 5.00$\times 10^{-8}$ A \\
\hline
$N$ & 1.08 \\
\hline
$R_S$ & 6 $\Omega$ \\
\hline
$V_B$ & 7 V \\
\hline
$C_j$ & 0.18 pF \\
\hline
$M$ & 0 \\
\hline
$P_B$ & 0.65 V \\
\hline
$E_G$ & 0.69 eV \\
\hline
$I_K$ & 0 A/m$^2$ \\
\hline
$I_{KR}$ & 0 A/m$^2$ \\
\hline
\end{tabular}
\caption{\label{tab:1}SPICE parameters of the Schottky diode model used for waveform-selective metasurfaces. }
\end{table}

\begin{table}[tb]
\centering
\begin{tabular}{|r|c|c|c|}
\hline
 & $h$ (mm) & $p$ (mm) & $g$ (mm) \\
\hline
280-MHz model & 15 & 260 & 10 \\
\hline
2.8-GHz model & 1.5 & 26 & 1 \\
\hline
28-GHz model & 0.13 & 2.7 & 0.1 \\
\hline
280-GHz model & 0.015 & 0.26 & 0.01 \\
\hline
\end{tabular}
\caption{\label{tab:2}Design parameters of waveform-selective metasurfaces. }
\end{table}

In this study we use the unit cell of the simple waveform-selective metasurface drawn in Fig.\ \ref{fig:1}. In this structure, the unit cell is composed of a square conducting patch, a dielectric substrate (Rogers3003) and a ground plane (perfect electric conductor: PEC), which resembles unit cells of existing metasurfaces \cite{EBGdevelopment}. However, the gap between conducting patches is connected by a set of four Schottky diodes that serve as a full-wave rectifier. Therefore, although the resonant frequency is determined by the same concept as the one that applies to ordinary metasurfaces \cite{EBGdevelopment}, the incoming waveform following a sine function is fully rectified to generate another waveform based on the modulus of the sine function. In this case, as seen in the Fourier expansion of the rectified waveform, the incoming frequency component is converted to an infinite set of components, although most of the energy is at zero frequency \cite{wakatsuchi2013waveform,wakatsuchi2015waveformSciRep,wakatsuchi2019waveform}. For this reason, the circuit configuration drawn in Fig.\ \ref{fig:1} enables exploitation of the transient phenomena well known in DC circuits even if the incoming waveform is an alternating current (AC) signal. In the particular case shown in Fig.\ \ref{fig:1}, the energy of a short pulse is temporarily stored in the capacitor and then dissipated with the parallel resistor, resulting in strong absorption. If the pulse width is sufficiently long, however, then the capacitor is fully charged such that electric charges induced by the incident wave cannot enter the internal circuit. Thus, the absorption performance is significantly reduced even at the same frequency. Other types of waveform-selective metasurfaces are further explained in the literature \cite{wakatsuchi2015waveformSciRep,vellucci2019waveform,wakatsuchi2019waveform}. In this study, we adopt a co-simulation method available from an ANSYS numerical simulator (Electronics Desktop 2020 R2) to simulate the abovementioned capacitor-based waveform-selective metasurface. Note that the Schottky diodes used in this study are modelled by referring to the SPICE parameters of a commercial product provided by Avago, specifically the HSMS-286x series (Table \ref{tab:1}). However, since we vary these SPICE parameters over a wide range, a similar conclusion is expected to be drawn in designing waveform-selective metasurfaces containing other diodes. 

\begin{figure}[tb]
\centering
\includegraphics[width=\linewidth]{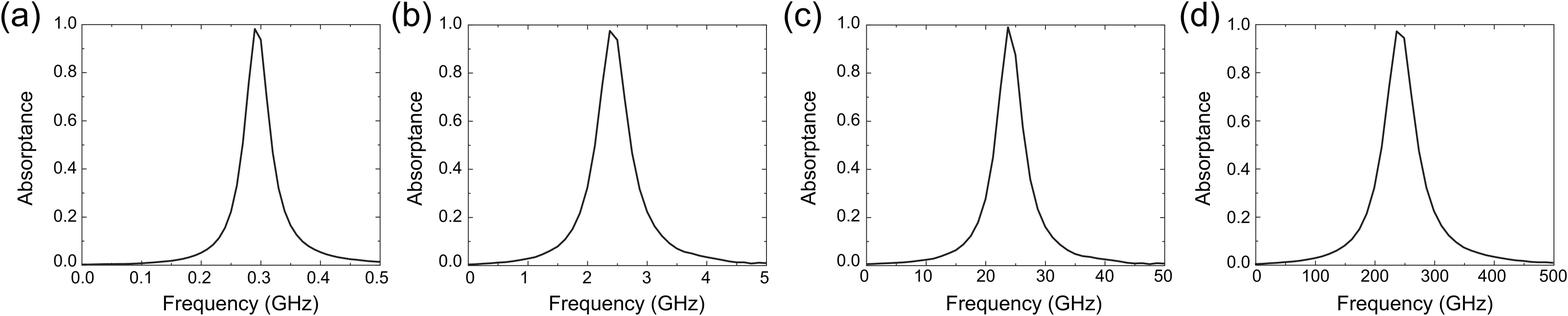}
\caption{Frequency-domain absorption profiles of simulation models using only an 120$\pi$-$\Omega$ resistor in the conductor gap without a diode bridge or a capacitor: (a) 280-MHz, (b) 2.8-GHz, (c) 28-GHz and (d) 280-GHz models.}
\label{fig:2}
%
\includegraphics[width=\linewidth]{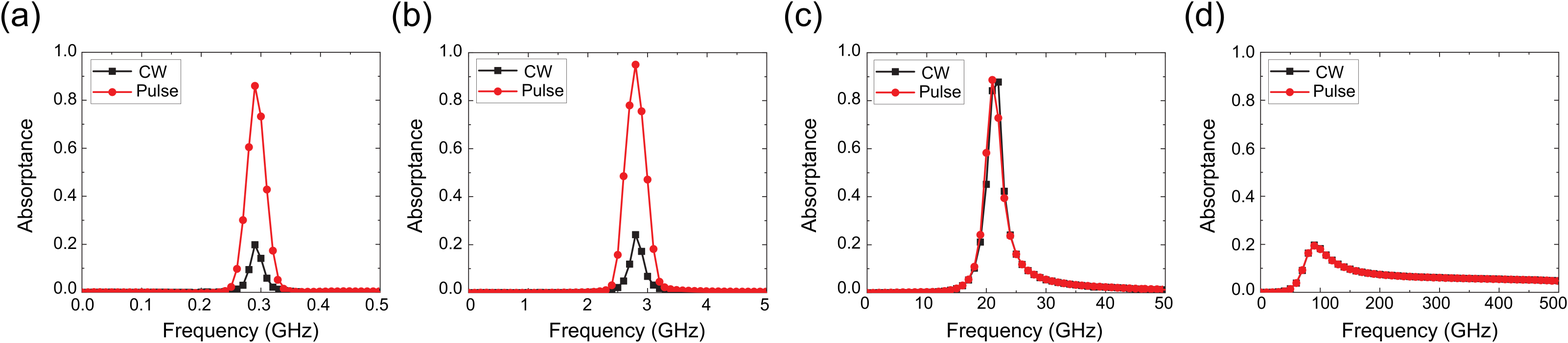}
\caption{Frequency-domain absorption profiles of simulation models using all the circuit components for high-power signals: (a) 280-MHz, (b) 2.8-GHz, (c) 28-GHz and (d) 280-GHz models. The input power was set to 0 dBm. }
\label{fig:3}
\end{figure}

First, we numerically studied how the operating frequency of the waveform-selective metasurface can be scaled to a smaller or larger frequency range. Here, we changed the physical design parameters to build simulation models near 280 MHz, 2.8 GHz, 28 GHz and 280 GHz (see the parameters in Table \ref{tab:2}). The substrate thickness was adjusted to be almost consistent with the thickness of existing printed circuit board (PCB, Rogers3003) products. Additionally, the incident pulse width was set to 500 ns, 50 ns, 5 ns and 0.5 ns for the 280-MHz, 2.8-GHz, 28-GHz and 280-GHz models, respectively, to maintain the relative bandwidth of the incident pulse (see the literature reporting that the spectrum spreads if the pulse width is too narrow \cite{wakatsuchi2015time}). Before evaluating the waveform-selective response, we simplified the waveform-selective metasurface models to replace all the circuit components with a single 120$\pi$-$\Omega$ resistor and investigate the operating frequency by using the abovementioned co-simulation method. The simulated absorptances are plotted in Fig.\ \ref{fig:2}, where the absorptance peak appeared at approximately 280 MHz, 2.8 GHz, 28 GHz and 280 GHz, as expected. 

\begin{figure}[tb]
\centering
\includegraphics[width=0.4\linewidth]{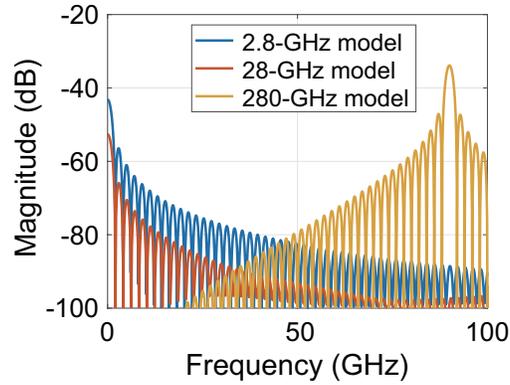}
\caption{Spectrum of the current at one of diodes used in the waveform-selective metasurface. The frequencies of the 2.8-GHz, 28-GHz and 280-GHz models were fixed at 2.8, 21 and 90 GHz, respectively, where the transmittance was most enhanced. The input power was set to 0 dBm. }
\label{fig:4}
\end{figure}

\begin{figure}[tb]
\centering
\includegraphics[width=0.4\linewidth]{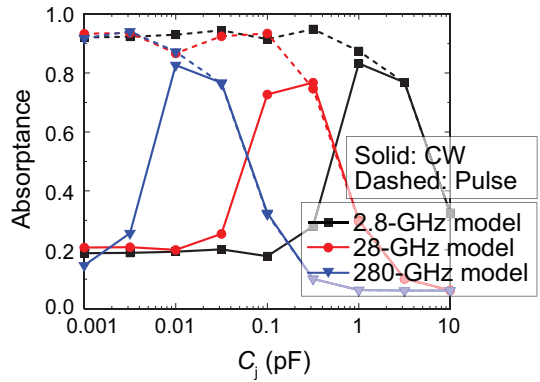}
\caption{Relationship between absorptance and $C_j$. The frequencies of the 2.8-GHz, 28-GHz and 280-GHz models were adjusted to maximize the transmittance. The input power was set to 0 dBm. }
\label{fig:5}
\end{figure}

After obtaining these results, we conducted a transient analysis to send sufficiently large signals (0 dBm) to the four models. The absorptance of the waveform-selective metasurface was obtained, as shown in Fig.\ \ref{fig:3}. According to this figure, the 280-MHz model and the 2.8-GHz model showed clear differences between the absorptance for short pulses and that for continuous waves (CWs) due to the abovementioned waveform-selective absorption mechanism. However, the 28-GHz model and the 280-GHz model exhibited almost no difference despite the different waveforms and the presence of the rectifying circuit. This occurred because the diodes contained a parasitic junction capacitance $C_j$, which short-circuited the gap between square patches in the high-frequency region. To readily understand these results, the current flowing into one of the four Schottky diodes was Fourier transformed, as plotted in Fig.\ \ref{fig:4}. This figure indicates that since the reactance of $C_j$ approached zero and the diodes no longer rectified induced electric charges, the largest frequency component appeared at a nonzero frequency in the 280-GHz model. Moreover, the $C_j$ used in SPICE parameters was varied, as shown in Fig.\ \ref{fig:5}, where the difference between the pulse absorptance and CW absorptance was found to increase if $C_j$ was sufficiently small. For instance, when $C_j=1$ fF, the 280-GHz model exhibited strong absorptance of approximately 0.9 for a short pulse and limited absorptance of less than 0.2 for a CW. Therefore, reducing $C_j$ is very important to distinguish a short pulse from a CW in the high-frequency range. 

Another point here is that the 280-GHz model had a very small gap of 0.015 mm between conducting patches. This indicates that the loaded circuit components, including diodes, must be smaller than 0.015 mm. From the viewpoint of semiconductor fabrication, small diode chips can be fabricated by using the stealth dicing technique if the chip dimensions are 0.1 mm or larger \cite{stealthDicing}. To further reduce the chip dimensions, directly fabricating conducting patterns of metasurfaces on chips, namely, not on PCBs, is more realistic \cite{10microFabrication,metaOnChip}. Additionally, $C_j$ can be reduced to the femtofarad order, as plotted in Fig.\ \ref{fig:5} \cite{10microFabrication}. However, this does not necessarily ensure that other diode characteristics such as optimal turn-on voltage and current are realized in this power range. Therefore, reaching a balance between $C_j$ and other parameters is important to optimize the performance of waveform-selective metasurfaces in the high-frequency range. 

\begin{figure}[tb]
\centering
\includegraphics[width=\linewidth]{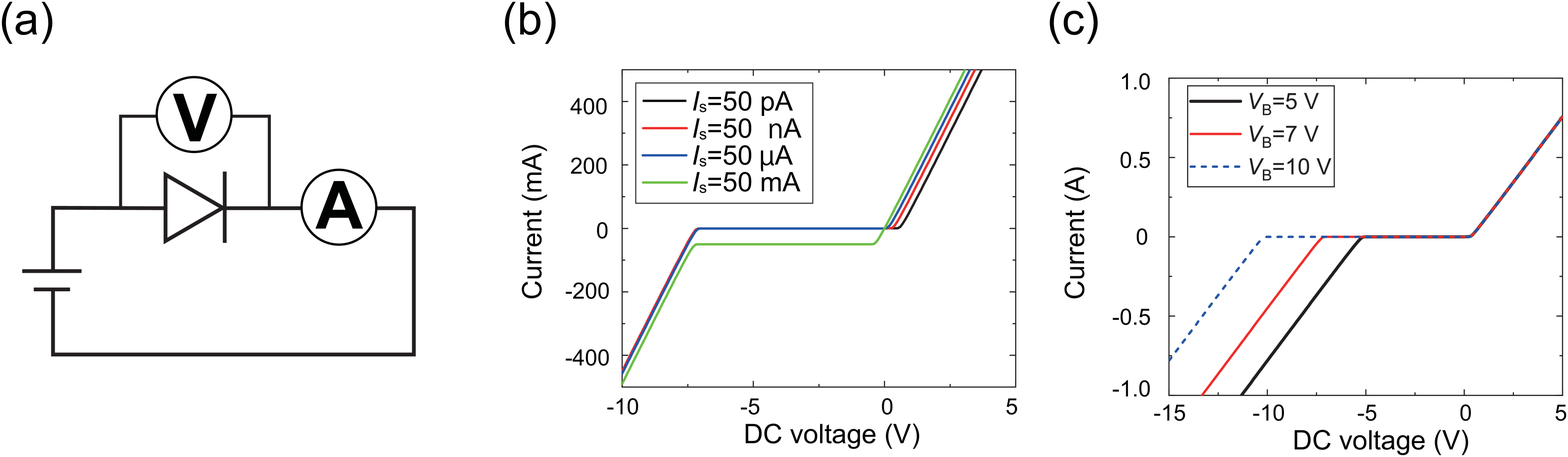}
\caption{Influence of saturation current $I_S$ and breakdown voltage $V_B$. (a) Circuit schematic. $IV$ curves with various (b) $I_S$s and (c) $V_B$s. }
\label{fig:6}
\end{figure}

Next, we investigated how the operating power level is determined by the SPICE parameters. To readily understand the relationship between the operating power level and diode properties, first, we plotted simple $IV$ curves of a single diode biased by a DC source, as drawn in Fig.\ \ref{fig:6}a, where we changed saturation current $I_S$ and breakdown voltage $V_{B}$. According to Fig.\ \ref{fig:6}b, the turn-on voltage was decreased by increasing $I_S$. At the same time, however, increasing $I_S$ allowed an increase in the backward current. Additionally, Fig.\ \ref{fig:6}c shows that $V_B$ simply determined the voltage to break down the diode. Specifically, the larger $V_B$ was, the larger the voltage that could be applied across the diode without current leakage.  

\begin{figure}[tb]
\centering
\includegraphics[width=0.8\linewidth]{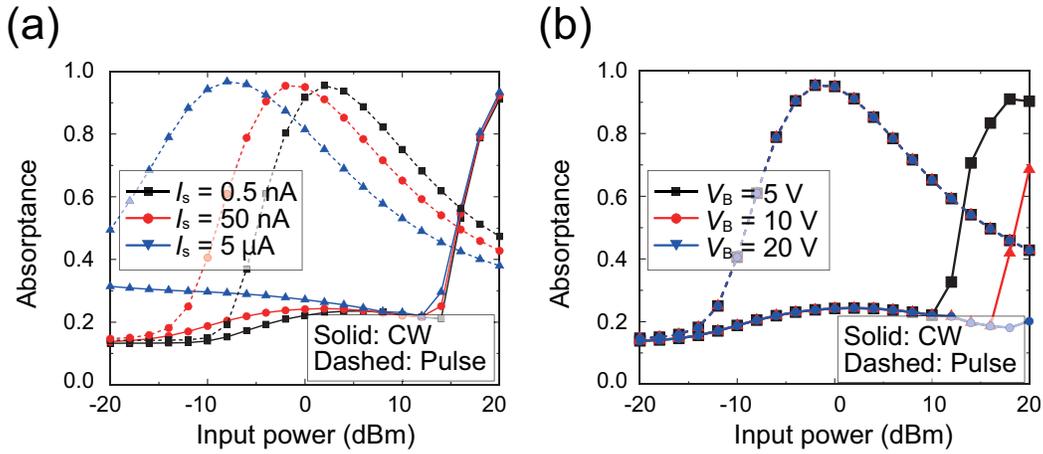}
\caption{Relationship between absorptance and $I_S$ or $V_B$ in the 2.8-GHz model. Absorptance with various (b) $I_S$s and (c) $V_B$s. The frequency and input power were fixed at 2.8 GHz and 0 dBm, respectively. }
\label{fig:7}
\end{figure}

Based on the simulation results in Fig.\ \ref{fig:6}, we then varied the $I_S$ and $V_B$ of the waveform-selective metasurface, as plotted in Fig.\ \ref{fig:7}. This figure shows that by increasing $I_S$, strong absorption was obtained for short pulses with a small power level (e.g., near -10 dBm with $I_S$ set to 0.5 nA). At the same time, however, the CW absorptance gradually increased since, as shown in Fig.\ \ref{fig:6}b, the diodes permitted more electric charge to come in from the backward direction as well. In addition, Fig.\ \ref{fig:7} indicates that by increasing $V_B$, the CW absorptance was suppressed for a wider range of the input power. This figure shows that if the CW absorptance starts increasing, then the input power exceeds the breakdown voltage of the diodes. We also varied other parameters such as forward knee current $I_K$ and backward knee current $I_{KR}$, each of which is related to the amount of current at the turn-on voltage and the breakdown voltage, respectively. Changes in $I_K$ and $I_{KR}$ also determine the waveform-selective response, as shown in Fig.\ \ref{fig:7}.

\begin{figure}[tb]
\centering
\includegraphics[width=0.8\linewidth]{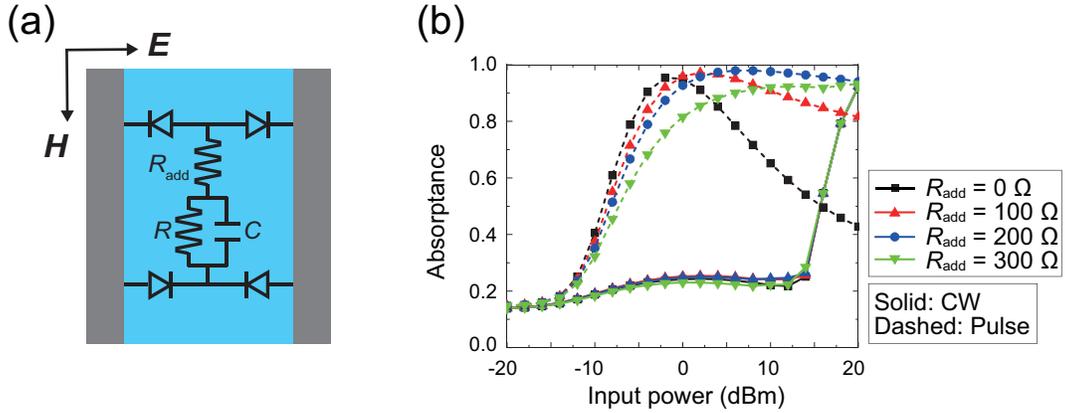}
\caption{Use of an additional series resistance $R_{add}$ to improve the dynamic range of the 2.8-GHz model. (a) Configuration of the circuit deployed between conductor edges. This circuit was deployed between conducting ptaches (see Fig.\ \ref{fig:1}a). (b) Absorptance with various $R_{add}$s. The frequency and input power were fixed at 2.8 GHz and 0 dBm, respectively.}
\label{fig:8}
\end{figure}

In Fig.\ \ref{fig:7}, we clarified how the operating power range was associated with $I_S$ and $V_B$. However, the difference between the pulse absorptance and the CW absorptance was maximized only in a narrow power range (e.g., around 0 dBm in Fig.\ \ref{fig:7}b). This occurred because this dynamic range was constrained by the change in the resistive component of the diodes \cite{aplEqCircuit4WSM}. To address this issue and minimize the influence of the resistive component of the diodes, one may add an additional series resistance $R_{add}$ to the inside of the diode bridge, as drawn in Fig.\ \ref{fig:8}a. In this case, the resistive component of the diodes still varies with the input power. However, the influence can be almost ignored if $R_{add}$ is moderately large. For this reason, the difference between the pulse absorptance and the CW absorptance was maintained at 0.7 or higher from 0 to 14 dBm when $R_{add}$ was set to 200 $\Omega$, as shown in Fig.\ \ref{fig:8}b. Thus, the dynamic range of the waveform selectivity can be readily broadened by adjusting $R_{add}$. 

\begin{figure}[tb]
\centering
\includegraphics[width=0.5\linewidth]{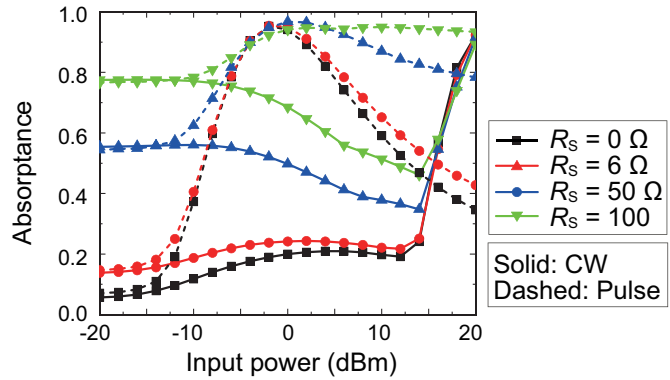}
\caption{Absorptance with various $R_S$s in the 2.8-GHz model. The frequency and input power were fixed at 2.8 GHz and 0 dBm, respectively.}
\label{fig:9}
\end{figure}

Moreover, the parasitic series resistive component of the diodes $R_S$ relates to the dynamic range of the waveform-selective metasurface to some extent. This is seen in Fig.\ \ref{fig:9}, where $R_S$ was changed from 0 to 100 $\Omega$. The absorptance for the short pulse was as large as 0.9 between 0 and 20 dBm when $R_S$ was 100 $\Omega$, which was similar to the pulse absorptance with $R_{add}$ = 200 $\Omega$ in Fig.\ \ref{fig:8}b. However, the absorptance for the CW was also increased in the same power range because $R_S$ was involved in not only the rectification process required for the waveform selectivity but also the intrinsic resonance mechanism of the patch structure (see the structure showing large absorptance in Fig.\ \ref{fig:2}). Therefore, including a resistive component inside the diode bridge as $R_{add}$ is important to increase the absorptance for short pulses but not CWs. 

\section*{Conclusion}
In this study, we investigated how the absorption profiles of waveform-selective metasurfaces are related to their SPICE parameters. First, we clarified the relationship between the SPICE parameters and the operating frequency. The operating frequencies of conventional metasurfaces that do not include circuit components are well known to be adjusted by simply varying the physical dimensions. In the case of waveform-selective metasurfaces, however, junction capacitance $C_j$ was shown to play an important role in the high-frequency range in maintaining the difference between the absorptance for a short pulse and that for a CW. Second, we showed how the operating power level of the waveform-selective metasurfaces is determined by the SPICE parameters. Our simulation results showed that saturation current $I_S$ and breakdown voltage $V_B$ are important for reducing the operating power level and maintaining small absorptance up to a large input power level, respectively. However, these two parameters were found to not contribute to broadening of the dynamic range of the waveform-selective metasurfaces, which can be improved by introducing an additional resistive component $R_{add}$ into the inside of the diode bridge. Thus, our study is expected to provide design guidelines for circuit-based waveform-selective metasurfaces to select/fabricate optimal diodes and enhance the waveform-selective performance at the target frequency and power level.


\providecommand{\noopsort}[1]{}\providecommand{\singleletter}[1]{#1}%

\section*{Acknowledgements}

This work was supported in part by the National Institute of Information and Communications Technology (NICT), Japan under the commissioned research No.\ 06201, the Japan Science and Technology Agency (JST) under the Precursory Research for Embryonic Science and Technology (PRESTO) No.\ JPMJPR193A and the Japanese Ministry of Internal Affairs and Communications (MIC) under the Strategic Information and Communications R$\&$D Promotion Program (SCOPE) No.\ 192106007.

\section*{Author contributions statement}

H.W. designed the entire project. S.I., H.H. and M.T. jointly conducted numerical simulations and analysed the results. H.W., A.A.F., C.Y., J.N, M.K., Y.O. and K.N. also considered the results. All authors reviewed the manuscript. 

\section*{Competing financial interests}
The authors declare no competing interests. 
\end{document}